\documentclass{article}
\usepackage{amsmath}
\usepackage{amssymb}
\newcommand{\be}{\begin{equation}}
\newcommand{\ee}{\end{equation}}

\begin{document}
\baselineskip18pt

\title{Frictionless Random Dynamics: Hydrodynamical Formalism}
\author{Rados\l aw Czopnik \\
 Institute of Theoretical Physics, University of Wroc\l aw,\\
 PL - 50 209 Wroc\l aw, Poland\\
 and\\
 Piotr Garbaczewski\thanks{E-mail: pgar@proton.if.uz.zgora.pl}\\
 Institute  of Physics, University of Zielona G\'{o}ra,\\
 PL - 65 069 Zielona G\'{o}ra, Poland}

\maketitle

\begin{abstract}
We investigate  an undamped random phase-space  dynamics in
deterministic external force fields (conservative and  magnetic
ones). By employing the hydrodynamical formalism for those
stochastic  processes we  analyze  microscopic  kinetic-type
"collision invariants" and their relationship to local
conservation laws (moment equations) in the fully nonequlibrium
context. We address an issue of the continual heat absorption
(particles "energization")  in the course of the process and  its
possible physical implementations.

\end{abstract}

PACS number(s): 02.50.-r, 05.40.-a, 03.65.-w

\section{Introduction}

The major topic discussed in the present paper will be  the
frictionless random   motion which is introduced in close
affinity with  second-order processes driven by white noise,
\cite{heinrichs,masoliver}.  In the course of motion an
unrestricted "energization" of particles is possible, a
phenomenon  which has not received much attention
 in the literature, although appears  to be set on  convincing
  phenomenological grounds in   specific (nontypical) physical
 surroundings, \cite{newman,newman1}.

A  classic problem  of an irreversible behaviour of a particle
embedded in the large encompassing system,
\cite{ford,uller,dorfman} pertains  to the  random evolution
of  tracer  particles in a fluid  (like  e.g. those performing
Brownian motion). In that case, a  particle - bath coupling is
expected  to imply quick  relaxation towards a
stationary  probability distribution (equilibrium Maxwellian
for a given a priori temperature $T$ of the surrounding fluid/bath)
 and thus making the
system amenable to standard fluctuation-dissipation theorems. The
dissipation of randomly acquired  energy \it back  to \rm the bath
is here  enforced by dynamical friction mechanisms,
\cite{lewis,chandra,nelson,dorfman}, although
 \it no \rm   balance between stochastic forcing (random
acceleration effects) and dissipative losses of energy is in fact
in existence in the course of the process, unless asymptotically,
\cite{uller,gar00}.

The  traditional picture of a tracer particle in a fluid/gas,
even if modeled in  terms of stochastic diffusion-type processes,
 is usually considered in  direct reference to the Boltzmann
 kinetic theory. That justifies an exploitation of properly
 defined microscopic collision rules
to yield dynamical friction (c.f. \cite{dorfman,liboff}) and then
to allow  for  a "derivation" (which breaks  down the microscopic
conservation laws  on the way, \cite{gar99}) of the Brownian
dynamics in suitable scaling limits. The crucial assumption about
a rapid decay  of correlation functions  amounts to  strong
friction and short relaxation time, \cite{dorfman,chandra,lewis}.

Let us mention  that, as reveals the hydrodynamical formalism of
the Brownian motion, \cite{klim},    Brownian particles "while
being thermalized",  show up a continual net absorption of
heat/energy from the reservoir  (that point is normally
disregarded in view of the very short relaxation times):
 so-called kinetic temperature  grows up to the actual
temperature of the bath, in parallel  with the mean kinetic
energy of the Brownian particle, \cite{gar00,czopnik,uller}, c.f.
also \cite{huang,schuss}. Indeed, \cite{schuss} chap.2, for $t>0$
we encounter on the average a continual  "heating"  phenomenon in
the course of which  an asymptotic equilibrium value $kT/m$ is
achieved: $E|\overrightarrow{u}|^2 = {\frac{q}\beta
}[1-\exp(-2\beta t)] \rightarrow \frac{kT}m $ where $q=\beta
kT/m$, while $\overrightarrow{u}$ stands for velocity random
variable. In this case the bath is  thought of as insensitive  to
this process and is supposed to remain in the perpetual state of
"statistical rest".

(Notice that such situation  is surely in conflict with dynamical
approaches to the thermalization of the Brownian particle,
\cite{jarzynski,gar99}, which take into account that the
particle-thermostat coupling acts both ways (consider reaction
forces, third Newton law in the mean etc.). In consequence,
 it is \it both  \rm the Brownian particle \it  and  \rm
the dissipating agent (bath, thermostat) which  need to thermalize
together i.e. simultaneously  evolve towards   a common state of
statistical equilibrium, cf. \cite{jarzynski}.)

 In below we shall consider purely classical prototype stochastic
 models which depart from most traditional Brownian motion in its
 Langevin (white noise), Kramers and Smoluchowski  versions.
 However our focus will be on an extremally nonequilibrium situation
 when no thermalisation is possible. Then, stationary (equilibrium)
state is definitely \it out \rm of  the reach, no
fluctuation-dissipation relationships are in existence and the
untamed  heating phenomenon is the  most conspicuous
characteristic of the randomly forced dynamics.

\subsection{Phenomenological motivations}

In contrast to the  well established frictional "Brownian motion
standards", we shall explore the \it opposite \rm regime of
dissipation - free theory. The departure from the standard theory
appears to be quite radical, since it refers to a different
physics, where dissipative time scales (if any) are \it much
longer \rm than the time duration of processes of interest,
including the particle life-time.

We are particularly motivated by two papers \cite{newman,newman1}
devoted to the thermal energization of particles which is due to
impulsive stochastic forcing. In fact, frictionless stochastic
processes were invoked to analyze situations present in
magnetospheric environments. Specifically, one deals there with
charged particles in a  locally uniform magnetic field  which
experience inetrmittent stochastic electrical forcing.

However, then a serious conceptual obstacle is present, if the
theory is considered without suitable reservations: one needs to
be able to control and eventually tame the kinetic energy growth,
since in the absence of friction we deal with an efficient
particle "energization" mechanism. That involves as well an
important issue (not addressed in the present paper) of suitable
thermalization/thermostatting mechanisms that would possibly tame
the otherwise unlimited "energization" of particles or make it
effectively irrelevant on the time-scales of interest.

Small random impulses occur within ambient fields in such way
that the smooth deterministic dynamics of a particle  is
irregularly  disrupted by sudden and very short bursts of
low-intensity noise. Without friction, we deal with an undamped
random walk  and   no fluctuation-dissipation relationship can be
established.

Phenomena of that kind are seldom mentioned in the traditional
 statistical physics research, since they may be realized in
rather unusual, from the point of view of the common practice,
environments.

Indeed, \cite{newman2} the physical nature of environments in
which the random propagation is investigated, appears to be
crucial.  In the original papers \cite{newman,newman1}, the
environment of interest was an astrophysical or space plasma,
where densities and therefore dissipation   are often reduced by
many orders of magnitude. That amounts to  dissipation time
scales which are much longer than those on which any random
forcing may take place. Moreover, physical processes  associated
with the thermalization were eliminated  in view of the fact that
particles entered geometrical regions of space where the plasma
instabilities (producing stochastic electric fields fluctuations)
and thus the source of randomness had disappeared. In contrast to
typical laboratory environments (e.g. plasmas), the high  spatial
heterogeneity present in space plasmas  makes random forcing to
occur only within very limited regions  of the non-stationary
environment and \it not \rm  to occur everywhere and for  all
times.

That means that it is the very nature of the environment which
causes the source of energization to disappear and reappear
spontaneously, while typical random bath models heavily rely on
the noise  spatial  homogeneity and  its "eternal" existence,
\cite{lewis}.

\subsection{Hydrodynamical formalism for random dynamics}

To elucidate the formal roots of our strategy and reasons for
studying the frictionless random dynamics by means of the
hy
drodynamical formalism, let us call our attention back to the
standard Ornstein-Uhlenbeck (dissipative) framework. The Langevin
equation for  mass $m$ particle in an external field of force (we
consider a conservative case)
$\overrightarrow{F}=\overrightarrow{F}\left(
\overrightarrow{x}\right) =-\overrightarrow {\nabla } V$ reads:

\begin{eqnarray}
\frac{d\overrightarrow{x}}{dt} &=&\overrightarrow{u} \\
\frac{d\overrightarrow{u}}{dt} &=&-\beta \overrightarrow{u}+\frac{%
\overrightarrow{F}}{m}+\overrightarrow{A}\left( t\right)
\end{eqnarray}

Random acceleration $\overrightarrow{A}\left( t\right) $ obeys
the white noise statistics:
 $\left\langle A_{i}\left( s\right) \right\rangle =$ $0$ and
 $\left\langle A_{i}\left( s\right) A_{j}\left( s^{\shortmid }\right)
\right\rangle =2q\delta \left( s-s^{\shortmid }\right) \delta _{ij}$,
where $i=1,2,3$.

Since things are specialized to the \it standard \rm Brownian
motion, we know  a priori that  noise intensity is determined by a
parameter  $q=D\beta ^{2}$  where  $D=\frac{kT}{m\beta }$, while the
friction parameter $\beta $  is given  by the Stokes formula $
m\beta =6\pi \eta a$ (or its analogue in case of the Lorentz gas,
\cite{dorfman}). Consequently, the effect of the surrounding
medium on the motion of the particle is described by two
parameters: friction constant $\beta $ and bath temperature $T$.
Assumptions about the asymptotic (equilibrium) Maxwell-Boltzmann
distribution and the fluid reaction upon the moving particle are here
implicit, \cite{chandra}.

The resulting (Markov)  phase - space diffusion  process is
completely determined by the transition probability density
$P\left(
\left. \overrightarrow{x},\overrightarrow{u},t\right|
\overrightarrow{x}_{0},%
\overrightarrow{u}_{0},t_{0}\right) $, which is typically expected
to be a fundamental solution of the Kramers equation:

\begin{equation}
\frac{\partial P}{\partial t}+\overrightarrow{u}\nabla _{\overrightarrow{x}%
}P+\left[ -\beta \overrightarrow{u}+\frac{\overrightarrow{F}}{m}\right]
\nabla _{\overrightarrow{u}}P=q\nabla _{\overrightarrow{u}}^{2}P
\end{equation}

The associated spatial Smoluchowski diffusion process  with a forward drift
$\overrightarrow{b}(\overrightarrow{x}) =
\frac{\overrightarrow{F}}{m\beta }$
 is analyzed in terms of  increments of the  normalized Wiener process
 $\overrightarrow{W}(t)$. The infinitesimal increment of the
 configuration (position) random variable $\overrightarrow{X}(t)$ reads:

\begin{equation}
d\overrightarrow{X}\left( t\right) =\frac{\overrightarrow{F}}{\ m\beta }dt+%
\sqrt{2D}d\overrightarrow{W}\left( t\right)
\end{equation}

The related Fokker-Planck equation for the spatial probability
density $\rho(\overrightarrow{x},t)$ reads $\partial _t\rho =
D\triangle \rho -  \overrightarrow{\nabla }\cdot (\rho
\overrightarrow{b})$ and explicitly employs the large friction
regime, \cite{nelson,chandra,gar99}.
 In fact, we take for granted that both time and space scales \it
of interest \rm (i.e. those upon which the accumulation of
relevant random events prove to be significant) largely exceed the
relaxation time interval $\beta ^{-1}$ and that dominant
contributions  "of interest" come
 from velocities $|\overrightarrow{u}|\leq (kT/m) = (q/\beta )^{1/2}$
 and that the corresponding variation of $\overrightarrow{r}$ is
 sufficiently small (actually it is of the order
 $|\overrightarrow{u}|/\beta \equiv (q/\beta ^3)^{1/2})$, \cite{chandra}.

Under those assumptions the Fokker -Planck equation for the
spatial Markov process arises as the  scaling ($\beta \gg 1$)
limit of the $0$-th order moment equation  associated with the
original Kramers law of random phase-space dynamics. In fact, by
following the traditional pattern of hydrodynamical formalism,
\cite{klim,huang}, we infer the closed system of two (which is
special to Markovian diffusions !) local conservation laws for the
Smoluchowski process, \cite{gar00,gar99}:
\begin{eqnarray}
\partial _{t}\rho + \overrightarrow{\nabla }\cdot \left(
\overrightarrow{v} \rho \right)%
 &=&0 \\
( \partial _{t} + \overrightarrow{v}
\cdot \overrightarrow{\nabla }) \overrightarrow{v}%
 &=& \overrightarrow{\nabla }
\left( \Omega -Q\right) .
\end{eqnarray}

Here (we use  a short-hand notation
$\overrightarrow{v}(\overrightarrow{x},t) \doteq
\overrightarrow{v}$)
\begin{equation}
\overrightarrow{v}(\overrightarrow{x},t) = \frac{%
\overrightarrow{F}}{\ m\beta }- D\frac{\overrightarrow{\nabla }
\rho }{\rho }
\end{equation}
 defines
 so-called current velocity of Brownian particles and,
 when inserted to Eq. (5),  transforms the continuity equation into the
 Fokker-Planck equation, \cite{nelson}.

Eq. (6) stands for the scaling limit of the first order moment
equation derivable from the kinetic equation (3) and directly
corresponds to the familiar Euler equation, characterizing the
momentum conservation law in  the lowest order approximation of
kinetic theory based on the Boltzmann equation, \cite{huang}.

However, the large friction regime enforces here a marked
difference in the local momentum conservation law, in comparison
with the standard  Euler equation for a  nonviscous fluid or gas.
Namely, instead of  the kinetic theory motivated expression for
e.g. rarified gas:

\begin{equation}
 (\partial _t + \overrightarrow{v} \cdot
\overrightarrow{\nabla }) \overrightarrow{v}  =
\frac{\overrightarrow{F}}{\ m}\,  -  \,
\frac{\overrightarrow{\nabla } P} \rho
 \end{equation}
where $P(\overrightarrow{x})$ stands for the   pressure function
(to be fixed by a suitable equation of state) and
$\overrightarrow{F}$ is the very same (conservative
$-\overrightarrow{\nabla  }V$) force acting upon particles  as
that appearing in the   Kramers equation (3), the Smoluchowski
regime (6) employs the emergent  volume force (notice the
positive sign) $ + \overrightarrow{\nabla }\Omega $ instead of
$-\overrightarrow{\nabla }V$:

\begin{equation}
\Omega =\frac{1}{2}\left( \frac{\overrightarrow{F}}{\ m\beta
}\right) ^{2}+D\overrightarrow{\nabla }\cdot \left(
\frac{\overrightarrow{F}}{\ m\beta }\right)
\end{equation}

and the pressure-type contribution $ - \overrightarrow{\nabla }Q$
where, \cite{gar99} (see also \cite{geilikman,guth})

\begin{equation}
Q=2D^{2}\frac{\Delta \rho ^{1/2}}{\rho ^{1/2}}
\end{equation}
and $\Delta = \overrightarrow{\nabla }^2$ is the Laplace
operator, does not leave any room for additional constraints upon
the system (like e.g. the familiar equation of state).

 To have a glimpse of a dramatic  difference between  physical
 messages conveyed respectively  by equations (6) and (8), it is
 enough to insert in (8) the standard equation of state
 $P(\overrightarrow{x}) = \alpha
\rho ^{\beta }$ with $\alpha , \beta > 0$ and choose
$\overrightarrow{F} = - \omega ^2 \overrightarrow{x}$ to
represent the harmonic attraction in Eqs. (2) - (9), see also
\cite{gar99}.

Markovian diffusion processes with the inverted sign of
$\overrightarrow{\nabla }(\Omega - Q)$ in the local momentum
conservation law (6) i. e. respecting

\begin{eqnarray}
( \partial _{t} + \overrightarrow{v}
\cdot \overrightarrow{\nabla }) \overrightarrow{v}%
 &=& \overrightarrow{\nabla }
\left( Q -\Omega \right)
\end{eqnarray}

instead of Eq. (6), were considered  in Ref. \cite{gar99}
as implementations of the "third Newton law in the mean".
Nonetheless, also under those premises,  the  volume force
term $-\overrightarrow{\nabla }\Omega$ in Eq.  (11)
does \it not \rm in general  coincide with the externally acting
conservative force contribution  (e.g. acceleration)
${\frac{1} m}\overrightarrow{F} = - {\frac{1} m}
\overrightarrow{\nabla } V$ akin to Eq. (8).

Accordingly,  the effects of  external force fields
acting upon particles  are significantly distorted while passing to the
local conservation laws in the large friction (Smoluchowski) regime.

That becomes   even more conspicuous  in case of the Brownian motion
of a charged  particle in the constant magnetic field,
\cite{czopnik,lewis,balescu}.  In the Smoluchowski  (large friction)
regime, friction completely smoothes out any rotational (due to the Lorentz force)
features of the process. In the corresponding local   momentum conservation law
there is \it no \rm volume force  contribution  at all and merely the
"pressure-type" potential $Q$ appears  in a  rescaled  form,
\cite{czopnik}:
\begin{equation}
Q = {\frac{\beta ^2}{\beta ^2 + \omega _c^2}} \cdot  2D^2
\frac{\triangle \rho ^{1/2}}{\rho ^{1/2}}
\end{equation}
where $\beta $ is the (large) friction parameter and $\omega _c =
\frac{q_eB}{mc}$ is the cyclotron  frequency of the charge $q_e$
particle in a constant homogeneous  magnetic field
$\overrightarrow{B}= (0,0,B)$. Clearly, for moderate  frequency
values $\omega _c$ (hence the magnetic field intensity) and
sufficiently   large $\beta $ even this minor scaling remnant of
the original Lorentz force would  effectively disappear,
yielding  Eq. (10).

This observation is to be compared  with results of  Refs.
\cite{newman,newman1} where,  in the absence of friction, the
rotational Lorentz force  input clearly  survives  when passing to
the local conservation laws, in plain contrast with the
Smoluchowski regime. By disregarding friction  (or considering it
to be  irrelevant on suitable time  scales)  it is  possible to
reproduce exactly the conservative external force acting upon
particles in the local conservation laws, \cite{gar91}.

Thus the  low noise regime,   if  combined with the sufficiently
short duration time of processes of interest (much lower than the
dissipation time scales)  sets the range of validity of local
conservation laws in the classic Euler form (8),  when demanded
to  comply with  the   stochastic diffusion-type microscopic
dynamics  assumption.

\section{Random motion of a free particle}

\subsection{R\'{e}sum\'{e}}

Let us consider the Langevin-type equation for a particle which is not suffering
any friction while being  subject to random accelerations
\begin{eqnarray}
\frac{dx}{dt} &=&u  \label{Lan1} \\
\frac{du}{dt} &=&A\left( t\right)  \notag
\end{eqnarray}

We fix initial conditions: $x\left( t_{0}\right) =x_{0}$,
$u\left( t_{0}\right) =u_{0}$, $t_{0}=0$. The fluctuating term
$A\left( t\right) $ is assumed to display a standard white noise
statistic: (i) $A\left( t\right) $ is independent of $u$, (ii)
$\left\langle A\left( s\right) \right \rangle =$
$%
0$ and $\left\langle A\left( s\right) A\left(
 s^{\shortmid }\right) \right\rangle =2q\delta \left( s-
 s^{\shortmid }\right) $, where $q$ is here left as an unspecified (albeit a priori
 physical) parameter.

This random dynamics induces a transition probability density
$P\left( \left. x,u,t\right| x_{0},u_{0},t_{0}\right) $ which
uniquely defines the  corresponding phase-space Markovian
diffusion process executed by $\left( x,u\right) $. Here, the
function $P\left( \left. x,u,t\right| x_{0},u_{0},t_{0}\right) $
is the fundamental solution of  the Fokker-Planck equation
associated with the above Langevin equation:

\begin{equation}
\frac{\partial P}{\partial t}=-u\frac{\partial P}{\partial x}+q\frac{%
\partial ^{2}P}{\partial u^{2}} \, . \label{Fokker1}
\end{equation}

A fundamental solution of Eq. (14) was  first given  by Kolmogorov
\cite{kolmogorov} and is a literature classic,
\cite{heinrichs,masoliver}. For the uniformity of  further
argumentation, we shall here  follow a direct solution method
\cite{wentzell,schuss} which employs properties of the involved
Wiener process $W_{t}$.

Since $\sqrt{2q}W_{t}=\int_{0}^{t}A\left( s\right) ds$, the
distribution of the random variable $S=u-u_{0}$ is gaussian with
mean zero and variance $2qt$. We define the displacement of the
particle as $x-x_{0}=\int_{0}^{t}u\left(
\eta \right) d\eta $. Hence, we have  $R=x-x_{0}-u_{0}t=\sqrt{2q}%
\int_{0}^{t}W_{s}ds$.

Because $S$ and$\ R$ are gaussian random variables, their joint probability
distribution $W\left( R,S\right) $\ is also gaussian and is given in terms
of mean values and covariances of $S$ and$\ R$. The mean values are equal to
zero, the covariance matrix is given that $C=\left(
\begin{array}{cc}
cov\left( R,R\right) & cov\left( R,S\right) \\
cov\left( S,R\right) & cov\left( S,S\right)
\end{array}
\right) =2q\left(
\begin{array}{cc}
\frac{1}{3}t^{3} & \frac{1}{2}t^{2} \\
\frac{1}{2}t^{2} & t
\end{array}
\right) $ and $cov\left( W_{s},W_{t}\right) =s\wedge t$.

Thus, $W\left( R,S\right) =\frac{1}{2\pi }\left( \frac{1}{\det C}\right) ^{%
\frac{1}{2}}\exp \left[ -\frac{1}{2}\left(
\begin{array}{cc}
R & S
\end{array}
\right) C^{-1}\left(
\begin{array}{c}
R \\
S
\end{array}
\right) \right] $

Since the problem of finding the transition probability distribution $%
P\left( \left. x,u,t\right| x_{0},u_{0},t_{0}\right) $ is
equivalent to finding the joint probability distribution $W\left(
R,S\right) $\ of random variables $S$ and $R$, we have thus
arrived at the following expression for (time homogeneous)
 $P\left( \left. x,u,t\right| x_{0},u_{0},t_{0}\right) $:
\begin{equation}
P\left( \left. x,u,t\right| x_{0},u_{0},t_{0}=0\right) =
\frac{1}{2\pi }\frac{%
\sqrt{12}}{2qt^{2}}\exp \left[ -\frac{\left( u-u_{0}\right) ^{2}}{4qt}-\frac{%
3\left( x-x_{0}-\frac{u+u_{0}}{2}t\right) ^{2}}{qt^{3}}\right]\, .
\label{kolm}
\end{equation}

We encounter a  situation when standard fluctuation-dissipation
relationships are manifestly invalid. Nonetheless, it is useful
to mention a useful computational connection with  the familiar
damped   case. Namely, let us consider the Langevin equation for
the free (would be Brownian) particle
 with the friction term
\begin{align}
\frac{dx}{dt}& =u  \label{eq1} \\
\frac{dv}{dt}& =-\beta u+A\left( t\right)  \label{eq2}
\end{align}
but no a priori presumed (fluctuation-dissipation) relationship
between $\beta $ and $q$. We  again fix  the  initial conditions:
$x\left( t_{0}\right) =x_{0}$, $u\left( t_{0}\right) =u_{0}$,
$t_{0}=0$.

Except for considering $\beta $ and $q$ as independent
parameters, we can directly exploit the classic result
\cite{chandra} for the corresponding transition probability
density  $P\left( \left. x,u,t\right| x_{0},u_{0},t_{0}=0\right)
$\ solving the  Kramers equation (compare e.g. Eq. (3))

\begin{equation}
\frac{\partial P}{\partial t}+ u\frac{\partial P}{\partial x}=\beta \frac{%
\partial }{\partial u}\left( Pu\right) +q\frac{\partial ^{2}P}{
\partial u^{2}%
}
\end{equation}

Its  solution is known to  arise in the form of a
 joint probability density $W\left( R,S\right) $\ of
random variables $S=u-u_{0}e^{-\beta t}$ and $R=x-x_{0}-\beta
^{-1}\left( 1-e^{-\beta t}\right) u_{0}$

\begin{equation}
W\left( R,S\right) =\left( \frac{1}{4\pi ^{2}\left( fg-h^{2}\right) }\right)
^{\frac{1}{2}}\exp \left[ -\frac{gR^{2}-2hRS+fS^{2}}{2\left( fg-h^{2}\right)
}\right]
\end{equation}
where the auxiliary coefficients are given that $f=\frac{q}{\beta ^{3}}%
\left( 2\beta t-e^{-2\beta t}+4e^{-\beta t}-3\right) $, $g=\frac{q}{\beta }%
\left( 1-e^{-2\beta t}\right) $, $h=\frac{q}{\beta ^{2}}\left( 1-e^{-\beta
t}\right) ^{2}$.

Since we assume that the intensity of noise $q$ is \it
independent \rm  of  the friction parameter $\beta $, we can
safely take the limit of $\beta \rightarrow 0$ in  the joint
probability density $W\left( R,S\right) $. Then, in view of the
obvious limiting values for $S=v-v_{0}$, $R=x-x_{0}-v_{0}t$,
$f=\frac{2}{3}qt^{3}$, $g=2qt$, $h=qt^{2}$  we readily arrive at
the previous frictionless-case expression.

Let us stress that it is the assumption about the independence of parameters
 $q$ and  $\beta $  which takes us away
 from the standard theory of Brownian motion.
 We cannot any longer expect that thermal equilibrium  conditions
can be approached  by standard dissipation mechanisms. We also
recall that in case of  the "normal"  dissipative  free Brownian
theory \cite{chandra}, its  physical formulation  is based on the
requirement that the distribution of velocities   must be given by
  the stationary Maxwell-Boltzmann  probability density
  with variance $\frac{kT}{m}$\ as $t\rightarrow \infty $.
  Moreover, the traditional Einstein-Smoluchowski theory relates  the spatial
diffusion coefficient $D$  to $\beta $ by the  fluctuation - dissipation
relation $D=\frac{kT}{m\beta }$. That appears to be consistent with the Kramers
picture  of the phase-space  random dynamics only  if we  a priori assume
that $q=\frac{kT}{m}%
\beta $, \cite{chandra,nelson}. Then $q=D\beta ^2$ and large $\beta $
definitely implies large $q$.

\subsection{Kramers equation and local conservation laws}

The stationary (time homogeneous) transition probability density for the
diffusion process without friction in phase-space has the form  (15)
and is the fundamental solution to Kramers (Fokker-Planck type) equation (14).
We are interested in passing to a hydrodynamical picture, following the traditional
recipes \cite{klim,huang}. To that end we need to propagate certain initial
probability density  and investigate effects of the random dynamics. Let us
 choose most obvious  example of:

\begin{equation}
\rho _{0}\left( x,u\right) =\left( \frac{1}{2\pi a^{2}}\right) ^{\frac{1}{2}%
}\exp \left( -\frac{\left( x-x_{ini}\right) ^{2}}{2a^{2}}\right)
\left( \frac{1}{2\pi b^{2}}\right) ^{\frac{1}{2}}\exp \left(
-\frac{\left( u-u_{ini}\right) ^{2}}{2b^{2}}\right)\, .
\end{equation}
so that  at time $t$ we have $\rho \left( x,u,t\right) =\int
P\left( \left. x,u,t\right| x_{0},u_{0},t_{0}=0\right) \rho
_{0}\left( x_{0},u_{0}\right) dx_{0}du_{0}$.

Since $P\left( \left. x,u,t\right| x_{0},u_{0},t_{0}\right) $ is
the fundamental solution of the Kramers equation,  the joint
density $\rho \left( x,u,t\right) $ is also the solution and  can
be written in  the familiar, \cite{chandra},  form  of Eq. (19)
for $\rho (x,u,t) = W(R,S)$. However, in the present case
functional entries are adopted to the frictionless motion and
read as follows:  $S=u-u_{ini}$, $R=x-x_{ini}-u_{ini}t$ with
 $$f=a^{2}+b^{2}t^{2}+\frac{2%
}{3}qt^{3}$$
 \begin{equation}
 g=b^{2}+2qt
\end{equation}
$$h=b^{2}t+qt^{2}\, .$$

The marginals $\rho \left( x,t\right) =\int \rho \left(
x,u,t\right) du$ and $\rho \left( u,t\right) =\int \rho \left(
x,u,t\right) dx$\ are

\begin{equation}
\rho \left( x,t\right) =\left( \frac{1}{2\pi f}\right) ^{\frac{1}{2}}\exp
\left( -\frac{R^{2}}{2f}\right) =\left( \frac{1}{2\pi \left(
a^{2}+b^{2}t^{2}+\frac{2}{3}qt^{3}\right) }\right) ^{\frac{1}{2}}\exp \left(
-\frac{\left( x-x_{ini}-u_{ini}t\right) ^{2}}{2\left( a^{2}+b^{2}t^{2}+\frac{%
2}{3}qt^{3}\right) }\right)
\end{equation}
and

\begin{equation}
\rho \left( u,t\right) =\left( \frac{1}{2\pi g}\right)
^{\frac{1}{2}}\exp \left( -\frac{S^{2}}{2g}\right) =\left(
\frac{1}{2\pi \left( b^{2}+2qt\right) }\right) ^{\frac{1}{2}}\exp
\left( -\frac{\left( u-u_{ini}\right) ^{2}}{2\left(
b^{2}+2qt\right) }\right)
\end{equation}

Let us introduce an auxiliary (reduced) distribution
$\widetilde{W}\left( S|R\right) =\frac{W\left( S,R\right) }{\int
W\left( S,R\right) dS}$ where in the denominator we recognize
the marginal spatial distribution $\int W\left( S,R\right) dS
\doteq w $ of $W(S,R)$.

Following the standard hydrodynamical picture method
\cite{klim,huang,gar00} we define local (configuration space
conditioned) moments: $\left\langle u\right\rangle _{x}=\int
u\widetilde{W}du$ and $\left\langle u^{2}\right\rangle _{x}=\int
u^{2}\widetilde{W}du$. First,  we compute

\begin{equation}
\left\langle u\right\rangle _{x} =  u_{ini}+\frac{h}{f}R
=u_{ini}+\frac{b^{2}t+qt^{2}}{a^{2}+b^{2}t^{2}+\frac{2}{3}qt^{3}}\left[
x-x_{ini}-u_{ini}t\right]
\end{equation}

\begin{equation}
\left\langle u^{2}\right\rangle _{x}-\left\langle u\right\rangle
_{x}^{2}=\left( g-\frac{h^{2}}{f}\right) =\frac{q\,t^{3}\,\left(
2\,b^{2}+q\,t\right) +3\,a^{2}\,\left( b^{2}+2\,q\,t\right) }{%
3\,a^{2}+t^{2}\,\left( 3\,b^{2}+2\,q\,t\right) }
\end{equation}

Now, the first two moment equations for the Kramers equation are
easily derivable. Namely, the continuity ($0$-th moment) and the
momentum conservation (first moment) equations come out in the
form

\begin{eqnarray}
\frac{\partial w}{\partial t}+\frac{\partial }{\partial x}\left(
\left\langle u\right\rangle _{x}w\right) &=&0 \\
\frac{\partial }{\partial t}\left( \left\langle u\right\rangle _{x}w\right) +%
\frac{\partial }{\partial x}\left( \left\langle u^{2}\right\rangle
_{x}w\right) &=&0
\end{eqnarray}

These equations yield the  local momentum conservation law in the
familiar form
\begin{equation}
\left( \frac{\partial }{\partial t}+\left\langle u\right\rangle _{x}\frac{%
\partial }{\partial x}\right) \left\langle u\right\rangle _{x}=-\frac{1}{w}%
\frac{\partial P_{kin}}{\partial x}  \label{lmcl_1}
\end{equation}
where we  encounter the standard \cite{huang} textbook  notion of
the pressure function
\begin{equation}
P_{kin}\left(
x,t\right) =\left[ \left\langle u^{2}\right\rangle _{x}-\left\langle
u\right\rangle _{x}^{2}\right] w\left( x,t\right) \, .
\end{equation}

The marginal density  $w$ obeys  $\frac{\nabla w}{w}=-2f\nabla %
\left[ \frac{\Delta w^{1/2}}{w^{1/2}}\right] $ and that in turn implies

\begin{equation}
-\frac{1}{w}\frac{\partial P_{kin}}{\partial x}=2\left(
fg-h^{2}\right) \nabla \left[ \frac{\Delta
w^{1/2}}{w^{1/2}}\right]\, .
\end{equation}

As a consequence, the local conservation law (\ref{lmcl_1}) takes
the form

\begin{equation}
\left( \frac{\partial }{\partial t}+\left\langle u\right\rangle
_{x}\nabla \right) \left\langle u\right\rangle _{x}=-\frac{\nabla
P_{kin}}{w}= + 2\left( fg-h^{2}\right) \nabla \left[ \frac{\Delta
w^{1/2}}{w^{1/2}}\right] \doteq + \nabla Q
\end{equation}
where (we point out the plus   sign in the above, see e.g.  Eq. (11))

\begin{equation}
fg-h^{2}=a^{2}b^{2}+2a^{2}qt+\frac{2}{3}b^{2}qt^{3}+
\frac{1}{3}q^{2}t^{4} \doteq D^2(t)
\end{equation}
and  by adopting the notation $D^2(t) \doteq  fg - h^2$  we get
$-{\frac{1}w} {\frac{\partial P_{kin}}{\partial x}} = + \nabla Q$
with the functional form of $Q$ given by Eq. 10. Here, instead of
a diffusion constant $D$ we insert  the (positive) time
-dependent function $D(t)$.

With those notational adjustments, we recognize in Eq. (31)  a
consistent Euler form of the local momentum conservation law,  in
case of vanishing volume forces (c.f. Eq. (8) and compare that
with Eq. (11) again).

 It is useful to mention that by employing Eq.(23) we can
readily evaluate the mean kinetic energy associated with the
frictionless free dynamics. Namely, there holds
\begin{equation}
<u(t)^2> - <u(t)>^2 = b^2 + 2qt
\end{equation}
where $<u(t)>= <u(0)> = u_{ini}$. That is to be compared
 with the  standard  Brownian motion expression,
 \cite{schuss}: $<u(t)^2> ={\frac{q}\beta }[1- \exp(-2\beta t)]$.
In the present case we definitely encounter an untamed,  continual
heating (energization of Refs. \cite{newman,newman1}) mechanism.

Let us observe that limits $a\rightarrow 0, b\rightarrow 0$ are
here under control and  allow to reproduce a consistent
hydrodynamical formalism for the  transition probability density
of the frictionless process. Indeed, in that case, we readily
reproduce observations of Refs. \cite{heinrichs,masoliver} that
$<u^2(t)> = 2qt$, while for the spatial mean displacement we
would get $<x^2(t)> = {\frac{2}3}qt^3$. The spatial process shows
features typical of an anomalous transport and obeys the
Fokker-Planck equation $\partial _t \rho = (qt^2) \Delta \rho $.

\section{Charged particle in a constant magnetic field}

Our analysis of a  free particle subject to random accelerations and no
friction can be extended to the case when  external deterministic force
fields are in action.  As
a specific example  we  shall discuss the dynamics of a  charged
particle in a constant magnetic field, see e.g. also
\cite{czopnik,balescu,taylor,kursunoglu}.

Magnetic field effect upon the charged particle in
practically  frictionless random environments  was  the main
objective of Refs. \cite{newman,newman1} in the formulation of
 a kinematic mechanism for particle energization. Those
 authors were primarily interested in the application of their model in
  concrete physical situations, hence they did not concentrate on finding
the general solution to the involved probabilistic problem.
They have evaluated  the mean square velocity and gave computer assisted
visualization of the  dynamics. However no general descriptions of
the related velocity space nor phase-space  stochastic process were given.
Thus no  information  was available about spatial features of the
dynamics, nor about the associated
macroscopic (hydrodynamical formalism) balance equations.

\subsection{Velocity-space process}

For a charged particle in a constant magnetic field and fluctuating
 electric environment, we consider the equation of motion in the form

\begin{equation}
\frac{d\overrightarrow{u}}{dt}=\frac{q_{e}}{mc}\overrightarrow{u}\times
\overrightarrow{B}+\overrightarrow{A}\left( t\right)  \label{Langevin}
\end{equation}

where $\overrightarrow{u}$ denotes the velocity of the particle of an
electric charge $q_{e}$ and mass $m$. A fluctuating part $\overrightarrow{A}%
\left( t\right) $ represents random forces of electric origin which are to obey
the standard white noise statistics assumptions (c.f. Section 2.1).

Let us assume for simplicity that magnetic field $\overrightarrow{B}$ is
directed along the z-axis of a Cartesian reference frame  so that
$\overrightarrow{B}=\left( 0,0,B\right) $ and $B=const$. In this case,
Eq. (\ref{Langevin}) takes the form

\begin{equation}
\frac{d\overrightarrow{u}}{dt}=-\Lambda \overrightarrow{u}+\overrightarrow{A}%
\left( t\right)  \label{LanII}
\end{equation}

where
$$\Lambda =\left(
\begin{array}{ccc}
0 & -\omega _{c} & 0 \\
\omega _{c} & 0 & 0 \\
0 & 0 & 0
\end{array}
\right) $$
 and $\omega _{c}=\frac{q_{e}B}{mc}$ denotes  the cyclotron frequency.

 The corresponding transition probability density $P\left( \overrightarrow{u},t|%
\overrightarrow{u}_{0},t_0=0\right) $ of the time-homogeneous
 Markov process   with the initial condition:
$P\left( \overrightarrow{%
u},t|\overrightarrow{u}_{0}\right) \rightarrow \delta ^{3}\left(
\overrightarrow{u}-\overrightarrow{u}_{0}\right) $ as $t\rightarrow 0$
has the  form:

\begin{equation}
P\left( \overrightarrow{u},t|\overrightarrow{u}_{0}\right) =\left( \frac{1}{%
4\pi qt}\right) ^{\frac{3}{2}}\exp \left( -\frac{\left| \overrightarrow{u}%
-U\left( t\right) \overrightarrow{u}_{0}\right| ^{2}}{4qt}\right)
\label{sol-2}
\end{equation}
where $U(t) = \exp(- \Lambda t)$, see e.g. Ref. \cite{czopnik}.

$P\left( \overrightarrow{u},t|\overrightarrow{u}_{0},t_0=0\right) $ is
the solution of the Fokker-Planck equation:

\begin{equation}
\frac{\partial P}{\partial t}=-\omega _{c}\left[ \nabla _{\overrightarrow{u}%
}\times P\overrightarrow{u}\right] _{i=3}+q\nabla _{\overrightarrow{u}}^{2}P
\label{F-P_1}
\end{equation}
where $\left[ \nabla _{\overrightarrow{u}}\times P\overrightarrow{u}\right]
_{i=3}=\frac{\partial }{\partial u_{1}}\left( Pu_{2}\right) -\frac{\partial
}{\partial u_{2}}\left( Pu_{1}\right) $.
The expression (\ref{sol-2}) can be readily extended to the case $t_{0}=s\neq
0$, $t\geq s$,

\begin{equation}
P\left( \overrightarrow{u},t|\overrightarrow{v},s\right) =\left( \frac{1}{%
4\pi q\left( t-s\right) }\right) ^{\frac{3}{2}}\exp \left( -\frac{\left|
\overrightarrow{u}-U\left( t-s\right) \overrightarrow{v}\right| ^{2}}{%
4q\left( t-s\right) }\right)\, .
\end{equation}

Let us consider in some detail the motion of an individual particle which
has an initial  velocity $\overrightarrow{v}_{0}$ at $t_{0}=0$, and
the  initial probability density $\varrho \left( \overrightarrow{v},s=0\right) =\varrho
_{0}\left( \overrightarrow{v}\right) =\delta \left( \overrightarrow{v}-%
\overrightarrow{v}_{0}\right) $. The propagation in time is controlled  by
the  transition density distribution $P\left( \overrightarrow{u},t|\overrightarrow{%
v},s\right) $ according to
\begin{eqnarray}
\varrho \left( \overrightarrow{u},t\right) &=&\int P\left( \overrightarrow{u}%
,t|\overrightarrow{v},s=0\right) \varrho _{0}\left( \overrightarrow{v}%
\right) dv \\
\varrho \left( \overrightarrow{u},t\right) &=&\left( \frac{1}{4\pi qt}%
\right) ^{\frac{3}{2}}\exp \left( -\frac{\left| \overrightarrow{u}-U\left(
t\right) \overrightarrow{u}_{0}\right| ^{2}}{4qt}\right)\, .
\end{eqnarray}

By evaluating mean values of the resultant density  for $i=1,2$
velocity components, we obtain:
\begin{eqnarray}
\left\langle u_{i}\left( t\right) \right\rangle &=&\int_{-\infty }^{\infty
}u_{i}\varrho \left( \overrightarrow{u},t\right) d\overrightarrow{u} \\
\left\langle \overrightarrow{u}\left( t\right) \right\rangle &=&\left(
\left\langle u_{1}\left( t\right) \right\rangle ,\left\langle u_{2}\left(
t\right) \right\rangle \right) =U\left( t\right) \overrightarrow{u}_{0} \\
\left\langle u_{i}^{2}\left( t\right) \right\rangle &=&\int_{-\infty
}^{\infty }u_{i}^{2}\varrho \left( \overrightarrow{u},t\right) d%
\overrightarrow{u} \\
\sigma ^{2}_i &=&\left\langle u_{i}^{2}\left( t\right) \right\rangle
-\left\langle u_{i}\left( t\right) \right\rangle ^{2}=2qt
\end{eqnarray}
we realize that the kinetic energy of the particle increases linearly with time: $%
\left\langle u^{2}\left( t\right) \right\rangle -\left\langle u^{2}\left(
0\right) \right\rangle =4qt$. This result  obviously coincides with less
explicit  calculations of \cite{newman}.

If we define the displacement $\overrightarrow{r}$ of the particle as $%
\overrightarrow{r}-\overrightarrow{r}_{0}=\int_{0}^{t}\overrightarrow{u}%
\left( \eta \right) d\eta $, we can extend the description of the process to
the phase space. An attempt to  specify the transition probability density
$P\left( \overrightarrow{r},\overrightarrow{u},t|\overrightarrow{r}_{0},%
\overrightarrow{u}_{0},t_{0}=0\right) $  is equivalent to finding
the joint distribution $W\left(
\overrightarrow{S,}\overrightarrow{R}\right) $\ of random vectors $%
\overrightarrow{S}$ and $\overrightarrow{R}$, which are defined by the
equations
\begin{eqnarray}
\overrightarrow{S} &=&\overrightarrow{u}-U\left( t\right) \overrightarrow{u}%
_{0}  \label{S_mag} \\
\overrightarrow{R} &=&\overrightarrow{r}-\overrightarrow{r}_{0}-\Lambda ^{-1}%
\left[ 1-U\left( t\right) \right] \overrightarrow{u}_{0}  \label{R_mag}
\end{eqnarray}

where $\Lambda ^{-1}=\left(
\begin{array}{cc}
0 & \omega _{c}^{-1} \\
-\omega _{c}^{-1} & 0
\end{array}
\right) $ and $U\left( t\right) =\left(
\begin{array}{cc}
\cos \omega _{c}t & \sin \omega _{c}t \\
-\sin \omega _{c}t & \cos \omega _{c}t
\end{array}
\right) $.

The joint probability density of  $\overrightarrow{S}$ and $%
\overrightarrow{R}$\ has the form

\begin{equation}
W\left( \overrightarrow{S},\overrightarrow{R}\right) =\frac{1}{4\pi
^{2}\left( fg-h^{2}-k^{2}\right) }\exp \left( -\frac{f\left| \overrightarrow{%
S}\right| ^{2}+g\left| \overrightarrow{R}\right| ^{2}-2h\overrightarrow{S}%
\cdot \overrightarrow{R}+2k\left( \overrightarrow{S}\times \overrightarrow{R}%
\right) _{i=3}}{2\left( fg-h^{2}-k^{2}\right) }\right)  \label{sol_mag}
\end{equation}
with   the auxiliary coefficients:
$$g=2qt$$
\begin{equation}
f=4q\frac{1}{\omega
_{c}^{2}}\left( t-\frac{1}{\omega _{c}}\sin \omega _{c}t\right)
\end{equation}
$$h=\frac{%
2q}{\omega _{c}^{2}}\left( 1-\cos \omega _{c}t\right) $$
$$k=-\frac{2q}{%
\omega _{c}}\left[ t-\frac{1}{\omega _{c}}\sin \omega _{c}t\right]\, .
$$

Eq. (\ref{sol_mag}) is the solution to the Fokker-Planck equation in the
phase space:

\begin{equation}
\frac{\partial W}{\partial t}+\overrightarrow{u}\nabla _{\overrightarrow{r}%
}W=-\omega _{c}\left[ \nabla _{\overrightarrow{u}}\times W\overrightarrow{u}%
\right] _{i=3}+q\nabla _{\overrightarrow{u}}^{2}W  \label{Kram_mag}
\end{equation}

where $\left[ \nabla _{\overrightarrow{u}}\times W\overrightarrow{u}\right]
_{i=3}=\frac{\partial }{\partial u_{1}}\left( Wu_{2}\right) -\frac{\partial
}{\partial u_{2}}\left( Wu_{1}\right) $.

\subsection{Kramers equation and local conservation laws}

The formula (\ref{sol_mag})  is an explicit  solution to the
Kramers Fokker-Planck equation  and coincides with the phase-space
 transition probability density
$P\left( \left.
\overrightarrow{r},\overrightarrow{u},t\right| \overrightarrow{r}_{0},%
\overrightarrow{u}_{0},t_{0}=0\right) $.

Let us consider the following  initial  phase-space probability
density
\begin{equation}
\rho _{0}\left( \overrightarrow{r}_{0},\overrightarrow{u}_{0}\right) =\frac{1%
}{2\pi a^{2}}\exp \left( -\frac{\left( \overrightarrow{r}_{0}-%
\overrightarrow{r}_{ini}\right) ^{2}}{2a^{2}}\right) \frac{1}{2\pi b^{2}}%
\exp \left( -\frac{\left( \overrightarrow{u}_{0}-\overrightarrow{u}%
_{ini}\right) ^{2}}{2b^{2}}\right)
\end{equation}
which is propagated in the course of the stochastic process so that
at time $t$ we have $\rho \left( \overrightarrow{r},\overrightarrow{u}%
,t\right) =\int P\left( \left. \overrightarrow{r},\overrightarrow{u}%
,t\right| \overrightarrow{r}_{0},\overrightarrow{u}_{0},t_{0}=0\right) \rho
_{0}\left( \overrightarrow{r}_{0},\overrightarrow{u}_{0}\right) d%
\overrightarrow{r}_{0}d\overrightarrow{u}_{0}$

\bigskip Since $P\left( \left. \overrightarrow{r},\overrightarrow{u}%
,t\right| \overrightarrow{r}_{0},\overrightarrow{u}_{0},t_{0}=0\right) $ is
the fundamental solution of the Kramers equation  the joint density $\rho \left(
\overrightarrow{r},\overrightarrow{u},t\right) $ is also the solution. This
joint density can be put into the form

\begin{equation}
\rho \left( \overrightarrow{r},\overrightarrow{u},t\right) =W\left(
\overrightarrow{S},\overrightarrow{R}\right) =
\end{equation}
$$
\frac{1}{4\pi ^{2}\left(
fg-h^{2}-k^{2}\right) }\exp \left( -\frac{f\left| \overrightarrow{S}\right|
^{2}+g\left| \overrightarrow{R}\right| ^{2}-2h\overrightarrow{S}\cdot
\overrightarrow{R}+2k\left( \overrightarrow{S}\times \overrightarrow{R}%
\right) _{i=3}}{2\left( fg-h^{2}-k^{2}\right) }\right)
$$
where
\begin{eqnarray}
\overrightarrow{S} &=&\overrightarrow{u}-U\left( t\right) \overrightarrow{u}%
_{ini} \\
\overrightarrow{R} &=&\overrightarrow{r}-\overrightarrow{r}_{ini}-\Lambda
^{-1}\left[ 1-U\left( t\right) \right] \overrightarrow{u}_{ini} \\
g &=&b^{2}+2qt \\
f &=&\frac{2b^{2}\omega _{c}\left( 1-\cos \omega _{c}t\right) +a^{2}\omega
_{c}^{3}+4q\left( \omega _{c}t-\sin \omega _{c}t\right) }{\omega _{c}^{3}} \\
h &=&\frac{2q\left( 1-\cos \omega _{c}t\right) +b^{2}\omega _{c}\sin \omega
_{c}t}{\omega _{c}^{2}} \\
k &=&-\frac{2q\left( \omega _{c}t-\sin \omega _{c}t\right) +b^{2}\omega
_{c}\left( 1-\cos \omega _{c}t\right) }{\omega _{c}^{2}}\, .
\end{eqnarray}

The marginals $\rho \left( \overrightarrow{r},t\right) =\int \rho \left(
\overrightarrow{r},\overrightarrow{u},t\right) d\overrightarrow{u}$ and $%
\rho \left( \overrightarrow{u},t\right) =\int \rho \left( \overrightarrow{r},%
\overrightarrow{u},t\right) d\overrightarrow{r}$\ read
\begin{equation}
\rho \left( \overrightarrow{r},t\right) =\frac{1}{2\pi f}\exp \left( -%
\frac{\left| \overrightarrow{R}\right| ^{2}}{2f}\right) =
\end{equation}
$$
\frac{1}{2\pi \frac{1}{\omega _{c}^{3}}\left( 2b^{2}\omega _{c}\left(
1-\cos \omega _{c}t\right) +a^{2}\omega _{c}^{3}+4q\left( \omega _{c}t-\sin
\omega _{c}t\right) \right) }
$$
$$
\times \exp \left( -\frac{\left| \overrightarrow{r}-\overrightarrow{r}%
_{ini}-\Lambda ^{-1}\left[ 1-U\left( t\right) \right] \overrightarrow{u}%
_{ini}\right| ^{2}}{2\frac{1}{\omega _{c}^{3}}\left( 2b^{2}\omega _{c}\left(
1-\cos \omega _{c}t\right) +a^{2}\omega _{c}^{3}+4q\left( \omega _{c}t-\sin
\omega _{c}t\right) \right) }\right)
$$
and
\begin{equation}
\rho \left( \overrightarrow{u},t\right) =\frac{1}{2\pi g}\exp \left( -%
\frac{\left| \overrightarrow{S}\right| ^{2}}{2g}\right) =
\frac{1}{2\pi \left( b^{2}+2qt\right) }\exp \left( -\frac{\left|
\overrightarrow{u}-U\left( t\right) \overrightarrow{u}_{ini}\right| ^{2}}{%
2\left( b^{2}+2qt\right) }\right)\, .
\end{equation}

Notice that in view of $<u_i^2(t)> - <u_i(t)>^2= b^2 + 2qt$,
where $<u_i(t)> = [U(t)\overrightarrow{u}_{ini}]_i$ with $i=1,2$, and
$<\overrightarrow{u}(t)>^2 = |U(t)\overrightarrow{u}_{ini}|^2=
|\overrightarrow{u}_{ini}|^2=<\overrightarrow{u}(0)>$, we have the same as
before (see e.g. Section 2) indicative of the linear in time energy growth:
\begin{equation}
<\overrightarrow{u}^2(t)>  - <\overrightarrow{u}(0)>^2 = 2b^2 + 4qt \, .
\end{equation}

We shall perform major steps of the hydrodynamical  analysis
following the pattern of Sec. 2. To this end  we introduce the
auxiliary distribution $\widetilde{W}\left( \overrightarrow{S}|%
\overrightarrow{R}\right) =\frac{W\left( \overrightarrow{S},\overrightarrow{R%
}\right) }{\int W\left( \overrightarrow{S},\overrightarrow{R}\right) d%
\overrightarrow{S}}$ which has an explicit form
\begin{equation}
\widetilde{W}\left( \overrightarrow{S}|\overrightarrow{R}\right) =\frac{1}{%
2\pi \frac{1}{f}\left( fg-h^{2}-k^{2}\right) }\exp \left( -\frac{\left|
\overrightarrow{S}-\overrightarrow{m}\right| ^{2}}{2\frac{1}{f}\left(
fg-h^{2}-k^{2}\right) }\right)  \label{aux_mag}
\end{equation}
where $\overrightarrow{m}=\frac{1}{f}\left(
hR_{1}-kR_{2},hR_{2}+kR_{1}\right) $, $\overrightarrow{R}$ and $%
\overrightarrow{S}$ are given by equations (52) and (53)
respectively.

We define local moments as $\left\langle u_{i}\right\rangle _{%
\overrightarrow{R}}=\int u_{i}\widetilde{W}d\overrightarrow{u}$ and $%
\left\langle u_{i}^{2}\right\rangle _{\overrightarrow{R}}=\int u_{i}^{2}%
\widetilde{W}d\overrightarrow{u}$ where $i=1,2$. It is apparent from the Eq.
(\ref{aux_mag}) that $\left\langle \overrightarrow{u}\right\rangle _{%
\overrightarrow{R}}=\left( \left\langle u_{1}\right\rangle _{\overrightarrow{%
R}},\left\langle u_{2}\right\rangle _{\overrightarrow{R}}\right) =U\left(
t\right) \overrightarrow{u}_{ini}+\overrightarrow{m}$ and $\left\langle
u_{i}^{2}\right\rangle _{\overrightarrow{R}}=\left\langle u_{i}\right\rangle
_{\overrightarrow{R}}^{2}+\frac{1}{f}\left( fg-h^{2}-k^{2}\right) $.

The first two moment equations for the Kramers equation (\ref{Kram_mag}) are
easily derivable:
\begin{eqnarray}
\partial _{t}w+ \overrightarrow{\nabla }\left[ \left\langle \overrightarrow{u}\right\rangle _{%
\overrightarrow{R}}w\right] &=&0 \\
\partial _{t}\left[ \left\langle u_{1}\right\rangle _{\overrightarrow{R}}w%
\right] +\frac{\partial }{\partial r_{1}}\left[ \left\langle
u_{1}^{2}\right\rangle _{\overrightarrow{R}}w\right] +\frac{\partial }{%
\partial r_{2}}\left[ \left\langle u_{1}\right\rangle _{\overrightarrow{R}%
}\left\langle u_{2}\right\rangle _{\overrightarrow{R}}w\right] &=&\omega
_{c}\left\langle u_{2}\right\rangle _{\overrightarrow{R}}w \\
\partial _{t}\left[ \left\langle u_{2}\right\rangle _{\overrightarrow{R}}w%
\right] +\frac{\partial }{\partial r_{2}}\left[ \left\langle
u_{2}^{2}\right\rangle _{\overrightarrow{R}}w\right] +\frac{\partial }{%
\partial r_{1}}\left[ \left\langle u_{1}\right\rangle _{\overrightarrow{R}%
}\left\langle u_{2}\right\rangle _{\overrightarrow{R}}w\right] &=&-\omega
_{c}\left\langle u_{1}\right\rangle _{\overrightarrow{R}}w
\end{eqnarray}

Those equations imply  the local momentum conservation law
\begin{eqnarray}
\left[ \partial _{t}+\left\langle \overrightarrow{u}\right\rangle _{%
\overrightarrow{R}} \overrightarrow{\nabla }\right] \left\langle \overrightarrow{u}%
\right\rangle _{\overrightarrow{R}} &=&-\Lambda \left\langle \overrightarrow{%
u}\right\rangle _{\overrightarrow{R}}-\frac{1}{w}\overrightarrow{\nabla} \cdot
\overleftrightarrow{P}  \label{lmcl_mag} \\
&=&\frac{q_{e}}{mc}\left\langle \overrightarrow{u}\right\rangle _{%
\overrightarrow{R}}\times \overrightarrow{B}-\frac{1}{w}
\overrightarrow{\nabla }\cdot
\overleftrightarrow{P}  \notag
\end{eqnarray}
where $\overrightarrow{\nabla }\cdot \overleftrightarrow{P}$ is a vector which i-th component
is equal to $\sum_{j}\frac{\partial P_{ij}}{\partial r_{j}},i,j=1,2$ and $%
\overleftrightarrow{P}$ denotes the tensor consisting of components $%
P_{ij}=\left\langle \left( u_{i}-\left\langle u_{i}\right\rangle _{%
\overrightarrow{R}}\right) \left( u_{j}-\left\langle u_{j}\right\rangle _{%
\overrightarrow{R}}\right) \right\rangle _{\overrightarrow{R}}w$, they are
given that $P_{11}=\left( \left\langle u_{1}^{2}\right\rangle _{%
\overrightarrow{R}}-\left\langle u_{1}\right\rangle _{\overrightarrow{R}%
}^{2}\right) w$, $P_{22}=\left( \left\langle u_{2}^{2}\right\rangle _{%
\overrightarrow{R}}-\left\langle u_{2}\right\rangle _{\overrightarrow{R}%
}^{2}\right) w$ and $P_{12}=P_{21}=0$.

Let us point out that
\begin{equation}
\left\langle u_{1}^{2}\right\rangle _{\overrightarrow{R}%
}-\left\langle u_{1}\right\rangle _{\overrightarrow{R}}^{2}=\left\langle
u_{2}^{2}\right\rangle _{\overrightarrow{R}}-\left\langle u_{2}\right\rangle
_{\overrightarrow{R}}^{2}=g-\frac{h^{2}+k^{2}}{f}
\end{equation}
displays a linear growth in time  $\sim qt$ for large  $t$.

Owing to relation
\begin{equation}
\left( -\frac{1}{w}\frac{\partial P_{11}}{\partial x},-\frac{1}{w}\frac{%
\partial P_{22}}{\partial y}\right) =2\left( fg-h^{2}-k^{2}\right)
\overrightarrow{\nabla }
\left[ \frac{\Delta w^{1/2}}{w^{1/2}}\right]
\end{equation}
the local momentum conservation law (\ref{lmcl_mag}) takes the form
\begin{equation}
\left[ \partial _{t}+\left\langle \overrightarrow{u}\right\rangle _{%
\overrightarrow{R}}\overrightarrow{\nabla }\right] \left\langle \overrightarrow{u}%
\right\rangle _{\overrightarrow{R}}=\frac{q_{e}}{mc}\left\langle
\overrightarrow{u}\right\rangle _{\overrightarrow{R}}\times \overrightarrow{B%
}+2\left( fg-h^{2}-k^{2}\right) \overrightarrow{\nabla }\left[ \frac{\Delta w^{1/2}}{w^{1/2}}%
\right]  \label{lmcl_mag_2}
\end{equation}
where the crucial time-dependent coefficient reads
\begin{equation}
fg-h^{2}-k^{2}=a^{2}\,b^{2}\,-\frac{8\,q^{2}}{\omega _{c}^{4}}+\frac{%
4\,b^{2}\,q\,t}{\,\omega _{c}^{2}}+\frac{4\,q^{2}\,t^{2}}{\,\omega _{c}^{2}}%
+2\,a^{2}\,q\,t+\frac{8\,q^{2}}{\omega _{c}^{4}}\,\cos (t\,\omega _{c})-%
\frac{4\,b^{2}\,q}{\omega _{c}^{3}}\,\sin (t\,\omega _{c}) \, .
\end{equation}

In contrast to the free dynamics coefficient (32), presently we
need to maintain a proper balance between (small) noise intensity
$q$ and the (not too large)  frequency $\omega _c$ to guarantee
a  positivity of  (69), thus  allowing for a subsequent
redefinition in terms of $D^2(t)$, c.f. Eqs. (32), (32) for
comparison. If so, we can safely write (keep in mind that we
consider exclusively the planar motion):
\begin{equation}
2(fg -h^2 -k^2)\overrightarrow{\nabla }
\left[ \frac{\Delta w^{1/2}}{w^{1/2}}\right] = +\overrightarrow{\nabla } Q
\end{equation}

The important feature of this local conservation law is that on its
right-hand side appears the very same force which is present in the
Kramers equation (\ref{Kram_mag}), as opposed to the corresponding
conservation law in the  Smoluchowski
diffusion regime, \cite{gar99}.

\section{Harmonically bound particle}

\subsection{Velocity-space process}

For completeness, let us analyze  in detail the case
of harmonically attracting force.
We consider the one-dimensional harmonic oscillator with
circular frequency $%
\omega $. The equation of motion reads (for further notational convenience we
replace the previous $A(t)$ by $\xi (t)$):

\begin{eqnarray}
\frac{dx}{dt} &=&u  \label{1} \\
\frac{du}{dt} &=&-\omega ^{2}x+ \xi \left( t\right)  \notag
\end{eqnarray}

or

\begin{equation}
\frac{d}{dt}\left(
\begin{array}{c}
x \\
u
\end{array}
\right) =\left(
\begin{array}{cc}
0 & 1 \\
-\omega ^{2} & 0
\end{array}
\right) \left(
\begin{array}{c}
x \\
u
\end{array}
\right) +\left(
\begin{array}{c}
0 \\
\xi  \left( t\right)
\end{array}
\right)  \label{1-bis}
\end{equation}

and differs from the Langevin equation in the standard Ornstein-Uhlenbeck
theory in that we have no friction term $-\beta u$.

The formal solution to this equation is given in terms of the resolvent
operator $R\left( t,t_{0}\right) =e^{A\left( t-t_{0}\right) }$, where $%
A=\left(
\begin{array}{cc}
0 & 1 \\
-\omega ^{2} & 0
\end{array}
\right) $and has the form

\begin{equation}
\overrightarrow{x}\left( t\right) =R\left( t,t_{0}\right) \overrightarrow{x}%
_{0}+\int_{t_{0}}^{t}R\left( t,s\right) \overrightarrow{\xi }\left( s\right)
ds \, .
\end{equation}

For the sake of simplicity of the formulae we introduce the following
notation $\overrightarrow{x}\left( t\right) =\left( x\left( t\right)
,u\left( t\right) \right) $ and in our case $\overrightarrow{\xi }\left(
s\right) =\left( 0,\xi \left( s\right) \right) $.

The characteristic equation of the matrix $A$ is $\lambda ^{2}+\omega ^{2}=0$%
, with the eigenvalues $\lambda _{1}=i\omega $ and $\lambda _{2}=-i\omega $.
The corresponding eigenvectors are $x^{1}=\left(
\begin{array}{c}
1 \\
\lambda _{1}
\end{array}
\right) $ and $x^{2}=\left(
\begin{array}{c}
1 \\
\lambda _{2}
\end{array}
\right) $. In the base of eigenvectors $\left\{ x^{1},x^{2}\right\} $ matrix
$A$ has the diagonal form $A^{\prime }=\left(
\begin{array}{cc}
\lambda _{1} & 0 \\
0 & \lambda _{2}
\end{array}
\right) $.

Since $e^{A^{\prime }\left( t-t_{0}\right) }=\left(
\begin{array}{cc}
e^{\lambda _{1}\left( t-t_{0}\right) } & 0 \\
0 & e^{\lambda _{2}\left( t-t_{0}\right) }
\end{array}
\right) $ we can now easily evaluate the resolvent operator $R\left(
t,t_{0}\right) =e^{A\left( t-t_{0}\right) }=Be^{A^{\prime }\left(
t-t_{0}\right) }B^{-1}$, where $B$ $=\left(
\begin{array}{cc}
1 & 1 \\
\lambda _{1} & \lambda _{2}
\end{array}
\right) $ is the matrix related to the change of base and $B$ $^{-1}=\frac{1%
}{\lambda _{2}-\lambda _{1}}\left(
\begin{array}{cc}
\lambda _{2} & -1 \\
-\lambda _{1} & 1
\end{array}
\right) $.
Namely, we have:
\begin{equation}
R\left( t,t_{0}\right) =\frac{1}{\lambda _{2}-\lambda _{1}}\left(
\begin{array}{cc}
\lambda _{2}e^{\lambda _{1}\left( t-t_{0}\right) }-\lambda _{1}e^{\lambda
_{2}\left( t-t_{0}\right) } & -e^{\lambda _{1}\left( t-t_{0}\right)
}+e^{\lambda _{2}\left( t-t_{0}\right) } \\
\lambda _{1}\lambda _{2}e^{\lambda _{1}\left( t-t_{0}\right) }-\lambda
_{1}\lambda _{2}e^{\lambda _{2}\left( t-t_{0}\right) } & -\lambda
_{1}e^{\lambda _{1}\left( t-t_{0}\right) }+\lambda _{2}e^{\lambda _{2}\left(
t-t_{0}\right) }
\end{array}
\right)
\end{equation}
which can be explicitly written as follows:
\begin{equation}
R\left( t,t_{0}\right) =\left(
\begin{array}{cc}
\cos \omega \left( t-t_{0}\right) & \frac{1}{\omega }\sin \omega \left(
t-t_{0}\right) \\
-\omega \sin \omega \left( t-t_{0}\right) & \cos \omega \left( t-t_{0}\right)
\end{array}
\right)
\end{equation}

The formal solution to (\ref{1}) has the form
\begin{eqnarray}
x &=&x_{0}\cos \omega \left( t-t_{0}\right) +u_{0}\frac{1}{\omega }\sin
\omega \left( t-t_{0}\right) +\int_{t_{0}}^{t}\frac{1}{\omega }\sin \omega
\left( t-s\right) \xi \left( s\right) ds \\
u &=&-x_{0}\omega \sin \omega \left( t-t_{0}\right) +u_{0}\cos \omega \left(
t-t_{0}\right) +\int_{t_{0}}^{t}\cos \omega \left( t-s\right) \xi \left(
s\right) ds\, .
\end{eqnarray}

Let us introduce the following notation
\begin{eqnarray}
R &=&x-x_{0}\cos \omega \left( t-t_{0}\right) -u_{0}\frac{1}{\omega }\sin
\omega \left( t-t_{0}\right) =\int_{t_{0}}^{t}\psi \left( s\right) \xi
\left( s\right) ds \\
S &=&u+x_{0}\omega \sin \omega \left( t-t_{0}\right) -u_{0}\cos \omega
\left( t-t_{0}\right) =\int_{t_{0}}^{t}\varphi \left( s\right) \xi \left(
s\right) ds
\end{eqnarray}
where $\psi \left( s\right) =\frac{1}{\omega }\sin \omega \left( t-s\right) $
and $\varphi \left( s\right) =\cos \omega \left( t-s\right) $.

If we define furthermore:
\begin{eqnarray}
f &=&2q\int_{t_{0}}^{t}\psi ^{2}\left( s\right) ds=\frac{2q}{\omega ^{2}}%
\int_{t_{0}}^{t}\sin ^{2}\omega \left( t-s\right) ds=\frac{q}{\omega ^{2}}%
\left( t-t_{0}\right) \left[ 1-\frac{\sin 2\omega \left( t-t_{0}\right) }{%
2\omega \left( t-t_{0}\right) }\right] \\
g &=&2q\int_{t_{0}}^{t}\varphi ^{2}\left( s\right) ds=2q\int_{t_{0}}^{t}\cos
^{2}\omega \left( t-s\right) ds=q\left( t-t_{0}\right) \left[ 1+\frac{\sin
2\omega \left( t-t_{0}\right) }{2\omega \left( t-t_{0}\right) }\right] \\
h &=&2q\int_{t_{0}}^{t}\varphi \left( s\right) \psi \left( s\right) ds=\frac{%
2q}{\omega }\int_{t_{0}}^{t}\sin \omega \left( t-s\right) \cos \omega \left(
t-s\right) ds=\frac{q}{\omega ^{2}}\sin ^{2}\omega \left( t-t_{0}\right)
\end{eqnarray}
we end up with a function $W\left( R,S\right)$ of the form (20)
which can be read  out as a transition probability density for the Markovian
phase-space process (85), here denoted as
$P\left( \left. x,v,t\right| x_{0},v_{0},t_{0}\right) $,  which clearly
solves the appropriate Kramers-Fokker-Planck equation:
\begin{equation}
\frac{\partial P}{\partial t}+v\frac{\partial P}{\partial x}=\omega ^{2}x%
\frac{\partial P}{\partial v}+q\frac{\partial ^{2}P}{\partial v^{2}}\, .
\end{equation}

\subsection{Local conservation laws}

As an initial density we choose $\rho _{0}\left( x,u\right)$ of Eq. (21)
and propagate that according to $\rho \left( x,v,t\right) =\int P\left( \left.
x,u,t\right| x_{0},u_{0},t_{0}=0\right) \rho _{0}\left( x_{0},v_{0}\right)
dx_{0}du_{0}$.

Since $P\left( \left. x,u,t\right| x_{0},u_{0},t_{0}\right) $ is the
fundamental solution of the Kramers equation,  the joint density
$\rho \left( x,u,t\right) $
is also the solution. This joint density can be put into the form
 $\rho \left( x,u,t\right) =W\left( R,S\right)$, c.f. Eq. (96),
where
\begin{eqnarray}
S &=&u+x_{ini}\omega \sin \omega t-u_{ini}\cos \omega t \\
R &=&x-x_{ini}\cos \omega t-u_{ini}\frac{1}{\omega }\sin \omega t \\
f &=&\frac{\omega \left( b^{2}+2\,q\,t+a^{2}\,w^{2}\right) +\left( -\left(
b^{2}\,\omega \right) +a^{2}\,\omega ^{3}\right) \,\cos (2\,t\,\omega
)-q\,\sin (2\,t\,\omega )}{2\,\omega ^{3}} \\
g &=&\frac{\omega \,\left( b^{2}+2\,q\,t+a^{2}\,\omega ^{2}\right) +\omega
\,\left( b^{2}-a^{2}\,\omega ^{2}\right) \,\cos (2\,t\,\omega )+q\,\sin
(2\,t\,\omega )}{2\,\omega } \\
h &=&\frac{\sin (t\,\omega )\,\left( \omega \left( b^{2}-a^{2}\,\omega
^{2}\right) \,\cos (t\,\omega )+q\,\sin (t\,\omega )\right) }{\omega ^{2}}
\, .
\end{eqnarray}

The marginals $\rho \left( x,t\right) =\int \rho \left( x,u,t\right) du$ and
$\rho \left( u,t\right) =\int \rho \left( x,u,t\right) dx$\ read:
\begin{eqnarray*}
\rho \left( x,t\right) &=&\left( \frac{1}{2\pi f}\right) ^{\frac{1}{2}}\exp
\left( -\frac{R^{2}}{2f}\right) =  \\
&&\left( \frac{1}{2\pi \frac{1}{2\,\omega ^{3}}\left( \omega \left(
b^{2}+2\,q\,t+a^{2}\,w^{2}\right) +\left( -\left( b^{2}\,\omega \right)
+a^{2}\,\omega ^{3}\right) \,\cos (2\,t\,\omega )-q\,\sin (2\,t\,\omega
)\right) }\right) ^{\frac{1}{2}} \\
&&\times \exp \left( -\frac{\left( x-x_{ini}\cos \omega t-v_{ini}\frac{1}{%
\omega }\sin \omega t\right) ^{2}}{2\frac{1}{2\,\omega ^{3}}\left( \omega
\left( b^{2}+2\,q\,t+a^{2}\,w^{2}\right) +\left( -\left( b^{2}\,\omega
\right) +a^{2}\,\omega ^{3}\right) \,\cos (2\,t\,\omega )-q\,\sin
(2\,t\,\omega )\right) }\right)
\end{eqnarray*}
and
\begin{eqnarray*}
\rho \left( u,t\right) &=&\left( \frac{1}{2\pi g}\right) ^{\frac{1}{2}}\exp
\left( -\frac{S^{2}}{2g}\right) \\
&=&\left( \frac{1}{2\pi \frac{1}{2\omega }\left( \omega \,\left(
b^{2}+2\,q\,t+a^{2}\,\omega ^{2}\right) +\omega \,\left( b^{2}-a^{2}\,\omega
^{2}\right) \,\cos (2\,t\,\omega )+q\,\sin (2\,t\,\omega )\right) }\right) ^{%
\frac{1}{2}} \\
&&\times \exp \left( -\frac{\left( u+x_{ini}\omega \sin \omega t-u_{ini}\cos
\omega t\right) ^{2}}{2\frac{1}{2\omega }\left( \omega \,\left(
b^{2}+2\,q\,t+a^{2}\,\omega ^{2}\right) +\omega \,\left( b^{2}-a^{2}\,\omega
^{2}\right) \,\cos (2\,t\,\omega )+q\,\sin (2\,t\,\omega )\right) }\right)
\end{eqnarray*}

At this point we can evaluate $<u(t)>^2= [x_{ini} \omega \sin(\omega t) -
u_{ini} \cos(\omega  t)]^2$ and next:
\begin{equation}
<u^2(t)> - <u(t)>^2 =\frac{\omega (b^2 + 2qt +a^2\omega ^2) +
\omega (b^2 - a^2\omega ^2)\cos(2t\omega ) + q\sin(2t\omega )}{2\omega }
\end{equation}
to have clearly exemplified the kinetic energy growth (e.g. $<u^2(t)> -
<u(0)>^2$).  Notice that by taking the $\omega \rightarrow 0$ limit, we
recover the force-free dynamics result (35).

As before, let us  introduce the auxiliary distribution
$\widetilde{W}\left( S|R\right) =%
\frac{W\left( S,R\right) }{\int W\left( S,R\right) dS}$ which has the
 standard Gaussian form

\begin{equation}
\widetilde{W}\left( S|R\right) =\left( \frac{1}{2\pi \left( g-\frac{h^{2}}{f}%
\right) }\right) ^{\frac{1}{2}}\exp \left( -\frac{\left| S-\frac{h}{f}%
R\right| ^{2}}{2\left( g-\frac{h^{2}}{f}\right) }\right)  \label{aux_harm}
\end{equation}

We define local moments $\left\langle u\right\rangle
_{x}=\int u\widetilde{W}du$ and $\left\langle u^{2}\right\rangle _{x}=
\int
u^{2}\widetilde{W}du$ so that
\begin{eqnarray}
\left\langle u\right\rangle _{x} &=&-x_{ini}\omega \sin \omega t+
u_{ini}\cos
\omega t+\frac{h}{f}\left[ x-x_{ini}\cos \omega t-u_{ini}\frac{1}{\omega }%
\sin \omega t\right] \\
\left\langle u^{2}\right\rangle _{x} &=&\left\langle u\right\rangle
_{x}^{2}+\left( g-\frac{h^{2}}{f}\right)\, .
\end{eqnarray}

and  local conservation laws follow:
\begin{eqnarray}
\frac{\partial w}{\partial t}+\frac{\partial }{\partial x}\left(
\left\langle u\right\rangle _{x}w\right) &=&0 \\
\left( \frac{\partial }{\partial t}+\left\langle u\right\rangle _{x}\frac{%
\partial }{\partial x}\right) \left\langle u\right\rangle _{x} &=&-\omega
^{2}x-\left[ \left\langle u^{2}\right\rangle _{x}-\left\langle
u\right\rangle _{x}^{2}\right] \frac{1}{w}\frac{\partial w}{\partial x}\, .
\label{lmca_harm}
\end{eqnarray}

The local momentum conservation law (\ref{lmca_harm}) takes the form

\begin{equation}
\left( \frac{\partial }{\partial t}+\left\langle u\right\rangle _{x}\frac{%
\partial }{\partial x}\right) \left\langle u\right\rangle _{x}=-\omega
^{2}x+2\left( fg-h^{2}\right) \nabla \left[ \frac{\Delta w^{1/2}}{w^{1/2}}%
\right]
\end{equation}
where the time-dependent coefficient  reads:
\begin{equation}
fg-h^{2}=
\end{equation}
$$
\frac{-q^{2}+2\,b^{2}\,q\,t\,\omega ^{2}+2\,q^{2}\,t^{2}\,\omega
^{2}+2\,a^{2}\,b^{2}\,\omega ^{4}+2\,a^{2}\,q\,t\,\omega ^{4}+q^{2}\,\cos
(2\,t\,\omega )+q\,\omega \left( -b^{2}+a^{2}\,\omega ^{2}\right) \,\sin
(2\,t\,\omega )}{2\,\omega ^{4}}\, .
$$

Again, a proper balance between the (small) intensity parameter $q$ and the
frequency $\omega $ must be maintained to secure the positivity of (96),
 since then only the notation $D^2(t)$ can be consistently introduced and
 the pressure-type contribution $+\nabla Q$ explicitly identified in
 Eq. (95).

\section{Kinetic theory viewpoint: role of "collision invariants"}

Previously, we have directly identified the microscopic motion
scenario as the   phase-space stochastic process  to be followed
by an individual particle. The usage of the hydrodynamical picture
implicitly refers to  the  collective (ensemble)   "gas" picture
with its corresponding Boltzmann kinetic theory. Within the
Boltzmann framework,  Fokker-Planck-Kramers equations are derived
under various simplifying assumptions, \cite{liboff,dorfman},  and
effectively play the role of oversimplified kinetic equations,
with   the Boltzmann collision integral being  replaced by
suitable differentiable expressions.

In particular, let us notice that all hither to considered Kramers
equations have the general kinetic form:
\begin{equation}
\left( \partial _t + \overrightarrow{u} \cdot
\overrightarrow{\nabla }_{\overrightarrow{u}} +  {\frac{1}m}
\overrightarrow{F}\cdot
\overrightarrow{\nabla  }_{\overrightarrow{u}} \right) f = C(f)
\end{equation}
where  $f=f(\overrightarrow{x},\overrightarrow{u},t)$ is a sought
for phase-space probability density while $C(f)$ is a substitute
for the Boltzmann collision integral, $m$ being a mass parameter.

Would we have worked with the "normal" kinetic equation in the
Boltzmann form, then a standard method to produce  local
conservation laws is known to employ  the collision invariants
and heavily relies on an explicit form of $C(f)$ which must
yield:  $\int C(f) d^3u = 0$, $\int \overrightarrow{u} C(f)
 d^3u= \overrightarrow{0}$ and  in addition $\int \overrightarrow{u}^2 C(f)  d^3u =0$.
 That  implies respectively  mass, momentum and  energy conservation laws,
 as a consequence of the  existence of  microscopically conserved additive
 quantities,   \cite{liboff,huang}, c.f. also the very recent paper
  \cite{kaniadakis}.

A concrete form of the "collision integral" $C(f)$ while adopted to  the
considered before stochastic processes, can be easily deduced from the
corresponding  Fokker-Planck-Kramers equations.
For example, the frictional dynamics  involves $C(f)  =
\left( q\overrightarrow{\nabla }^2_{\overrightarrow{u}}
+ \beta \overrightarrow{u}
 \cdot \overrightarrow{\nabla }_{\overrightarrow{u}}\right) f$  while
 by omitting the $\beta $-dependent term we are left with the frictionless
 variant $C(f) = q \overrightarrow{\nabla }^2_{\overrightarrow{u}} f$.

It is important to notice that    only
$\int C(f)d^3u=0$ holds true  identically for  both   frictional and
frictionless  cases. The frictional case does not respect two other
kinetic identities.
On the other hand, the frictionless dynamics respects one more identity,
namely $\int  \overrightarrow{u} C(f)  d^3u= \overrightarrow{0}$, but \it
does \rm violate the  third one related  to the energy conservation rule.

As an almost trivial remark we may consider an observation that  one more step
in the hierarchy of gradually more constrained kinetic problems may be
completed  by choosing suitable solutions of the Liouville equation,
which we obtain by passing to the $q \rightarrow 0$ limit in Eqs. (14),
(37) and (83). That  corresponds to replacing the stochastically forced
dynamical systems by deterministic Newton laws, c.f. (13), (34), (71).
Since for such solutions, the right-hand-side of Eq. (97)  identically
vanishes, we obviously have $\int \overrightarrow{u}^2C(f)d^3u = 0$.

For those Liouville-type probability densities, all three collision
invariants  do vanish, satisfying the demands put forward in Ref.
\cite{kaniadakis}  to support the "idea that the classical kinetics
could be underlying quantum mechanics".  The point is that there is
manifestly \it no \rm collisional kinetics involved in this  particular
 case although all minimal requirements of Ref. \cite{kaniadakis} are
 fulfilled.

By performing $ q\rightarrow 0$ limits of  previously considered
joint velocity-space  probability densities   we pass smoothly
 to  deterministic solutions of the  respective Liouville
 equations in the  familiar form of the positive Wigner functions
  (up to notation adjustments,  see e.g. \cite{tak,hai}),
 with rather obvious hydrodynamical connotations.

 As  an illustrative example of this reasoning let us  reproduce in
 detail the  $q\rightarrow 0$ limit of  the frictionless dynamics of
 Section 2.  We have (tilde is used to discriminate between $q=0$
 and $q\neq 0$ cases):
\begin{equation}
\widetilde{\rho }\left( x,u,t\right) =\frac{1}{2\pi
\sqrt{a^{2}b^{2}}}\exp \left( -\frac{\left( u-u_{ini}\right)
^{2}}{2b^{2}}-\frac{\left( x-x_{ini}-tu\right)
^{2}}{2a^{2}}\right) \, .
\end{equation}

Here, the normalization is preserved, since  we have  $\int
\left( \int \widetilde{\rho }\left( x,u,t\right) dx\right)
du=\int \left( \int \widetilde{\rho }\left( x,u,t\right)
du\right) dx=1$, while  for the marginals
$\widetilde{\rho }\left( x,t\right) =\int \widetilde{\rho }%
\left( x,u,t\right) du$ and $\widetilde{\rho }\left( u,t\right)
=\int \widetilde{\rho }\left( x,u,t\right) dx$, in the very   same
$q \rightarrow  0$ limit we get:

\begin{equation}
\widetilde{\rho }\left( u,t\right) =\left( \frac{1}{2\pi b^{2}}\right) ^{%
\frac{1}{2}}\exp \left( -\frac{\left( u-u_{ini}\right)
^{2}}{2b^{2}}\right)
\end{equation}
and
\begin{equation}
\widetilde{\rho }\left( x,t\right) =\left( \frac{1}{2\pi \left(
a^{2}+b^{2}t^{2}\right) }\right) ^{\frac{1}{2}}\exp \left(
-\frac{\left( x-x_{ini}-u_{ini}t\right) ^{2}}{2\left(
a^{2}+b^{2}t^{2}\right) }\right) \, .
\end{equation}

In order to find local moments $\left\langle u\right\rangle _{x}=
\int u\frac{%
\widetilde{\rho }\left( x,u,t\right) }{\widetilde{\rho }\left( x,t\right) }%
du $ and $\left\langle u^{2}\right\rangle _{x}=\int u^{2}\frac{\widetilde{%
\rho }\left( x,u,t\right) }{\widetilde{\rho }\left( x,t\right)
}du$  of the joint probability density $\widetilde{\rho }(x,u,t)$
we compute

\begin{equation}
\frac{\widetilde{\rho }\left( x,u,t\right) }{\widetilde{\rho
}\left(
x,t\right) }=\left( \frac{a^{2}+b^{2}t^{2}}{2\pi a^{2}b^{2}}\right) ^{\frac{1%
}{2}}\exp \left( -\frac{\left( u-u_{ini}-\frac{b^{2}t}{a^{2}+b^{2}t^{2}}%
\left( x-x_{ini}-u_{ini}t\right) \right) ^{2}}{2\frac{a^{2}b^{2}}{%
a^{2}+b^{2}t^{2}}}\right)
\end{equation}

Consequently, the local moments are given by
\begin{equation}
\left\langle u\right\rangle _{x}=u_{ini}+\frac{b^{2}t}{a^{2}+b^{2}t^{2}}%
\left( x-x_{ini}-u_{ini}t\right)
\end{equation}

\begin{equation}
\left\langle u^{2}\right\rangle _{x}=\left\langle u\right\rangle _{x}^{2}+%
\frac{a^{2}b^{2}}{a^{2}+b^{2}t^{2}}
\end{equation}
and the  force-free Euler-type equation in the form (8), c.f.
\cite{gar00} is clearly recovered
\begin{equation}
\left( \frac{\partial }{\partial t}+\left\langle u\right\rangle
_{x}\nabla
\right) \left\langle u\right\rangle _{x}=2a^{2}b^{2}\nabla \left[ \frac{%
\Delta \widetilde{\rho }\left( x,t\right) ^{1/2}}{\widetilde{\rho
}\left( x,t\right) ^{1/2}}\right] \doteq  \nabla Q
\end{equation}
where (c.f. eq. (10)) we may identify $D^2=a^2b^2$ with a squared
diffusion constant, \cite{nelson,gar00}.
By setting formally $D= {\frac{\hbar }{2m}}$ we recover the standard
quantum mechanical "hydrodynamics", however with no intervention of
any nontrivial kinetics, contrary to the hypothesis of Ref.
\cite{kaniadakis}.

Quite analogously, the planar motion in a uniform magnetic field
gives rise to the Euler-type equation
\begin{equation}
\left[ \partial _{t}+\left\langle \overrightarrow{u}\right\rangle _{%
\overrightarrow{R}}\nabla \right] \left\langle \overrightarrow{u}%
\right\rangle _{\overrightarrow{R}}=\frac{q_{e}}{mc}\left\langle
\overrightarrow{u}\right\rangle _{\overrightarrow{R}}
\times \overrightarrow{B%
}+2a^{2}\,b^{2}\nabla \left[ \frac{\Delta {\widetilde{\rho }}^{1/2}}
{{\widetilde{\rho }}^{1/2}}\right]
\doteq  \frac{\overrightarrow{F}}m + \overrightarrow{\nabla }Q
\end{equation}
in conformity with the analogous result recovered in Eq. (69).
Here, $\overrightarrow{F}$ denotes the Lorentz force contribution.
The  harmonic case implies:
\begin{equation}
\left( \frac{\partial }{\partial t}+\left\langle u\right\rangle _{x}\frac{%
\partial }{\partial x}\right) \left\langle u\right\rangle _{x}=-\omega
^{2}x+2a^{2}\,b^{2}\nabla \left[ \frac{\Delta \widetilde{\rho }
^{1/2}}{\widetilde{\rho } ^{1/2}}\right]
\end{equation}
again with no   genuine kinetics involved.

 \section{Concluding remarks}

We have investigated in  detail exemplary solvable cases of the frictionless
random dynamics. All of them share a common feature  to yield the
local momentum conservation law in the specific form

\begin{equation}
\left[ \partial _{t}+\left\langle \overrightarrow{u}\right\rangle _{%
\overrightarrow{R}}\overrightarrow{\nabla }\right] \left\langle \overrightarrow{u}%
\right\rangle _{\overrightarrow{R}}= \frac{\overrightarrow{F}}m +2d(t)
\overrightarrow{\nabla }\left[ \frac{%
\Delta w^{1/2}}{w^{1/2}}\right]
\end{equation}

where $\overrightarrow{F}$ denotes external force  acting on
the particle, and $d=\left( \det C\right) ^{\frac{1}{n}}$ where $C$ is the
covariance matrix of random variables (vectors) $\overrightarrow{S}$ and $%
\overrightarrow{R}$ (defined for each system) and $n$ stands for the
dimension of configuration space of appropriate system.

\begin{enumerate}
\item  free particle: $\overrightarrow{R}=x$, $F\equiv 0$, $n=1$

\item  charged particle in a constant magnetic field: $\overrightarrow{R}%
=\left( x,y\right) $, $\overrightarrow{F}=\frac{q_{e}}{c}\left\langle
\overrightarrow{u}\right\rangle _{\overrightarrow{R}}\times \overrightarrow{B%
}$, $n=2$

\item  harmonically bound particle: $\overrightarrow{R}=x$, $F=- m \omega ^{2}x$%
, $n=1$
\end{enumerate}

In case of harmonic and magnetic confinement, we have identified  parameter
range regimes  that  allow for a positivity of the time dependent coefficient
 $d(t) \doteq D^2(t)$ (in the force-free case it is positive with no
 reservations), when the pressure-type contribution in Eq. (107) acquires
 a characteristic form of
 $- {\frac{\overrightarrow{\nabla } \cdot \overleftrightarrow{P}} w} =
 + \overrightarrow{\nabla }Q $.

 By investigating the frictionless diffusion scenario we have given
 support to the validity of Euler-type momentum conservation laws
  in a carefully defined stochastic process context.
In particular, that resolves problems with  a consistent
theoretical framework for diffusion processes which would show
up  Lorentz force effects in the local mean (e.g. rotational
dynamics) for long times (exceeding a single period, c.f
\cite{czopnik,gar97}).
However, the frictionless motion needs developing methods of control for the
otherwise untamed mean energy growth (energization phenomenon).

As a byproduct of the discussion we have demonstrated that the usage of three
 traditional "collision invariants" representing microscopically
 conserved additive quantities does not produce a sufficiently
 general kinetic theory background for the derivation of the probabilistic
 counterpart of the Schr\"{o}dinger picture quantum dynamics, at variance with
 the recent proposal of Ref. \cite{kaniadakis}.  The previous
 discussion indicates as well that neither dissipative nor
 non-dissipative stochastic phase-space processes  based on the
 white-noise kinetics are valid candidates to that end.
 \\

{\bf Acknowldegement:}
 We would like to thank Dr Karol  Pesz for pointing Refs.
 \cite{heinrichs,masoliver}  to our attention. We are also willing words of
 gratitude to Professor William I. Newman for inspiring correspondence.
 One of the authors (P. G.) receives support from the KBN research grant.

\end{document}